\documentclass{article}
\usepackage{spconf,amsmath,graphicx}
\usepackage{booktabs,adjustbox,makecell}
\usepackage{multirow,flushend}
\usepackage{amsfonts,mathtools,nccmath}
\usepackage{microtype}

\usepackage[labelformat=simple]{subcaption}

\usepackage[font=small,labelfont=bf]{caption}
\usepackage{bbm}
\usepackage{pifont}
\usepackage{enumitem}
\usepackage{xcolor}
\usepackage[linesnumbered,ruled,vlined]{algorithm2e}

\SetCommentSty{mycommfont}

\SetKwInput{KwInput}{Input}                
\SetKwInput{KwOutput}{Output}              
\SetKwInput{KwInit}{Init.}              

\usepackage{stfloats}
\fnbelowfloat 

\DeclarePairedDelimiter{\nint}\lfloor\rceil

\title{DOVER-Lap: A Method for Combining Overlap-aware Diarization Outputs}

\name{\begin{tabular}{c}Desh Raj$^1$, Leibny Paola Garcia-Perera$^{1,2}$, Zili Huang$^1$, Shinji Watanabe$^1$, \\
Daniel Povey$^3$, Andreas Stolcke$^4$, Sanjeev Khudanpur$^{1,2}$\thanks{This work was partially supported by grants from the JHU Applied Physics Laboratory, Nanyang Technological University, Hitachi Ltd., Japan, and the Government of Israel.}\end{tabular}}

\address{
  $^1$Center for Language and Speech Processing \& $^2$Human Language Technology Center of Excellence \\ 
  The Johns Hopkins University, Baltimore, MD 21218, USA\\
  $^3$Xiaomi Corp., Beijing, China, 
  $^4$Amazon Alexa Speech, Sunnyvale, CA, USA}
\email{\{draj2, lgarci27, shinjiw, khudanpur\}@jhu.edu, dpovey@gmail.com, stolcke@amazon.com}

\begin{document}

\maketitle
\begin{abstract}
Several advances have been made recently towards handling overlapping speech for speaker diarization. Since speech and natural language tasks often benefit from ensemble techniques, we propose an algorithm for combining outputs from such diarization systems through majority voting. Our method, DOVER-Lap, is inspired from the recently proposed DOVER algorithm, but is designed to handle overlapping segments in diarization outputs. We also modify the pair-wise incremental label mapping strategy used in DOVER, and propose an approximation algorithm based on weighted k-partite graph matching, which performs this mapping using a global cost tensor. We demonstrate the strength of our method by combining outputs from diverse systems --- clustering-based, region proposal networks, and target-speaker voice activity detection --- on AMI and LibriCSS datasets, where it consistently outperforms the single best system. Additionally, we show that DOVER-Lap can be used for late fusion in multichannel diarization, and compares favorably with early fusion methods like beamforming.
\end{abstract}

\section{Introduction}

Speaker diarization (or ``who spoke when?'') is the task of segmenting speech into homogeneous speaker-specific regions~\cite{Mir2012SpeakerDA, Tranter2006AnOO}. Until recently, conventional diarization systems~\cite{GarciaRomero2017SpeakerDU, Sun2018SpeakerDW} used a framework that involved clustering of fixed-dimensional embeddings, usually i-vectors~\cite{Dehak2011FrontEndFA}, d-vectors~\cite{Variani2014DeepNN}, or x-vectors~\cite{Snyder2018XVectorsRD}. Since these systems often used hard cluster assignments (via agglomerative or spectral clustering), they inherently assumed a ``single-speaker'' setting, i.e., they ignored overlaps completely. 

There have been efforts to solve the overlap problem in speaker diarization, with the existing approaches falling into two categories. In the first framework, a separate overlap detection module identifies segments which contain overlapping speech. This may be done using hidden Markov models (HMMs)~\cite{boakye2008overlapped, Huijbregts2009SpeechOD, Yella2012SpeakerDO} or using neural networks~\cite{Geiger2013DetectingOS, Andrei2017DetectingOS, Hagerer2017EnhancingLR, Kunesov2019DetectionOO}. Once overlaps are detected, they may be used to assign additional speaker labels to the overlapping segments. Recently, \cite{Bullock2019OverlapawareDR} proposed overlap-aware resegmentation, which leverages the variational Bayes (VB)-HMM method used originally for diarization in~\cite{Dez2018SpeakerDB}, and applied to resegmentation in~\cite{Sell2015DiarizationRI}. In the second framework, end-to-end systems such as EEND~\cite{Fujita2020EndtoEndND} or RPN~\cite{Huang2020SpeakerDW} are used to perform overlapping diarization in a supervised setting. A new diarization technique inspired by target speaker extraction methods has also been proposed recently~\cite{Medennikov2020TargetSpeakerVA}. With these advances, it has become possible to tackle overlapping speech for diarization.

However, these systems often have complementary strengths, and their combination may lead to better diarization performance, since machine learning tasks usually benefit from an ensemble of different systems~\cite{Rokach2009EnsemblebasedC}. ROVER~\cite{Fiscus1997APS} is a popular post-processing method for combining the outputs of speech recognition systems through weighted majority voting. This method was generalized to be used with n-best lists or lattices, and called confusion network combination (CNC)~\cite{evermann2000posterior,stolcke2000sri}. Lattice combination has also proved to be effective in systems developed for community challenges such as the CHiME-6 challenge~\cite{Arora2020TheJM}. However, it has traditionally been difficult to combine diarization systems at the output level due to the time-segmented nature of the hypotheses. Recently, DOVER~\cite{Stolcke2019DoverAM} was proposed as an algorithm to perform such combination through weighted majority voting (similar to ROVER for ASR) on homogeneous single-speaker regions across a recording. Evaluation results showed that it improves over random (or even oracle) channel selection.

While DOVER provides a convenient method to combine diarization outputs, it is limited since it cannot handle outputs containing overlapping segments. This limitation is of particular significance if we consider the increasing application of diarization to real multispeaker conversations containing high overlaps (such as the AMI~\cite{Carletta2005TheAM} meeting corpus and the CHiME challenges~\cite{Barker2018TheF,Watanabe2020CHiME6CT}), and the availability of methods that can perform diarization in such settings~\cite{Bullock2019OverlapawareDR,Huang2020SpeakerDW,Fujita2020EndtoEndND}. As such, it may be useful to devise an algorithm which can perform hypotheses combination in overlap-aware settings. We propose such a method in this paper, calling it DOVER-Lap (D\underline{OVER} + \underline{Over}lap) (DL). Our method builds upon DOVER in two aspects: (i) instead of pair-wise incremental label mapping, it uses an approximation algorithm that is globally informed by all pair-wise costs, and (ii) it can assign multiple speakers to a region based on the voting decision. We demonstrate through our experiments conducted on AMI and LibriCSS that DL can efficiently combine hypotheses from very different systems, ranging from conventional clustering-based to more recent end-to-end neural methods, while improving the DER performance over the single best system. To demonstrate the wide applicability of our technique, we also apply it to multichannel diarization through late fusion of outputs from far-field array microphones. We show that late fusion using DL is competitive with early fusion using dereverberation and beamforming, without any need for enhancement. Similar efforts have been made concurrently to extend DOVER for overlapping diarization hypotheses~\cite{Xiao2020MicrosoftSD}, but they rely on the same label mapping technique which is used in DOVER. Additionally, unlike our threshold-free label voting method, their modification uses a user-defined threshold per speaker, in order to assign speakers to each time region.

The key contributions in this paper are three-fold. First, we propose a fast and robust method for combining diarization hypothesis across multiple systems with overlap handling. An important part of this contribution is reformulating the label mapping using an approximation algorithm for the weighted maximal $k$-partite matching problem. Second, we demonstrate the effectiveness of our method on diverse systems, such as overlap-aware spectral clustering, VB resegmentation, region proposal network (RPN), and target-speaker voice activity detection (TS-VAD). To the best of our knowledge, a combination of hypotheses from such diverse systems has not been studied before. Finally, we show that our method can additionally be used to perform multichannel diarization on far-field array microphones through late fusion. The code for DOVER-Lap is made publicly available at: {\small \texttt{https://github.com/desh2608/dover-lap}}.

\section{The DOVER-Lap Algorithm}

\subsection{Problem formulation}
\label{sec:problem}

We assume that we are given the ``hypotheses'' $H_k$ from $K$ diarization systems, i.e., $H_1,\ldots,H_K$, where each system can perform overlapping speaker assignment and may contain different number of speakers, $N_1,\ldots,N_K$. Mathematically, $H_k$ can be written as a set of tuples containing time intervals and the corresponding speaker assignment. If $H_k$ has $N_k$ speakers $\{H_k^1,\ldots,H_k^{N_k}\}$, then 

\begin{equation}
    H_k = \{(\Delta_{k,\theta}, H_k^n): n \in \{1,\ldots,N_k\}\},
\end{equation}
where $\Delta_{k,\theta} = [t_{k,\theta},t_{k,\theta+1}]$ is a time interval, and $\theta$ is a time-stamp index.

We define the problem of combining the hypothesis as finding a joint diarization hypothesis $\hat{H} = f(H_1,\ldots,H_K)$ such that $\hat{H}$ minimizes the chosen diarization error metric with respect to an unknown reference. 

A straightforward approach for solving this problem involves dividing the input hypotheses into small time durations, and performing majority voting individually with each such region. However, there are two issues with this solution. First, to perform any kind of voting (within a region), the system outputs need to be in the same label space. Second, overlap-aware systems may contain different number of speakers within any such region, and for overlap-aware combination, the voting mechanism needs to account for this possibility. In the next sections, we describe how DOVER solves these problems, and how our DL algorithm improves upon the solution.

\begin{figure}[t]
\begin{subfigure}{0.49\linewidth}
\centering
\includegraphics[width=0.9\linewidth]{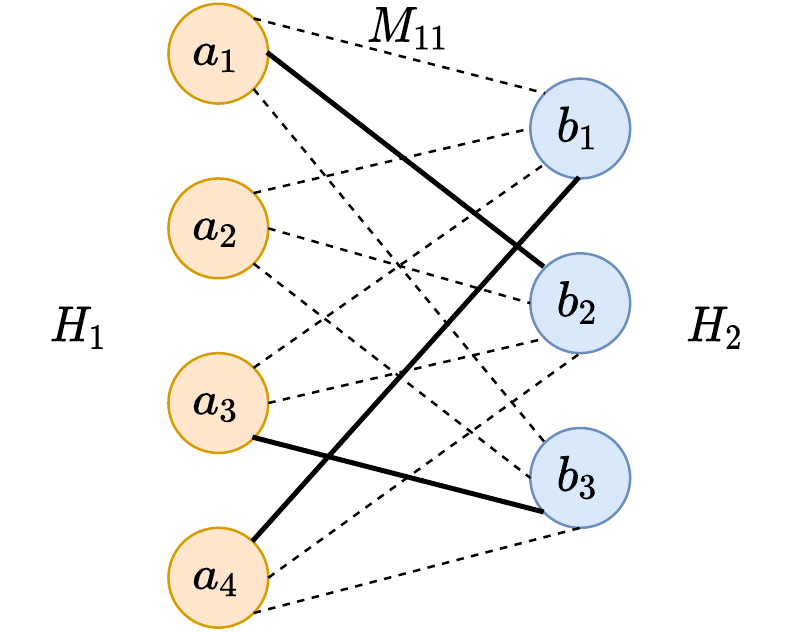}
\label{fig:bipartite_graph}
\end{subfigure}
\begin{subfigure}{0.49\linewidth}
\centering
\includegraphics[width=0.6\linewidth]{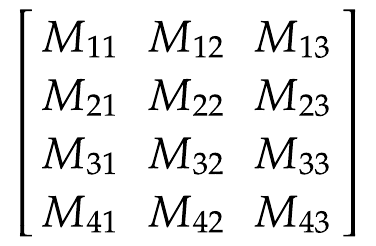}
\label{fig:bipartite_matrix}
\end{subfigure}\hfill
\caption{DOVER uses the Hungarian method pair-wise for label mapping. It iteratively solves the weighted bipartite graph matching problem using linear sum assignment on the cost matrix.}
\label{fig:hungarian}
\end{figure}

\vspace{-0.5em}
\subsection{Preliminary: DOVER}
\label{sec:dover}

DOVER (diarization output voting error reduction) is a weighted majority voting based method that combines diarization hypotheses~\cite{Stolcke2019DoverAM}. It comprises two stages: label mapping, and label voting, where each stage addresses one of the problems outlined in the previous section.

In the \textbf{label mapping} stage, the objective is to find a set of $N$ speaker labels $\{\hat{H}^1,\ldots,\hat{H}^N\}$ for our combined output $\hat{H}$, and a mapping function $\mathcal{S}$ that takes as input one of the speaker labels from the given hypotheses and maps it to a label in the combined output, i.e., $\mathcal{S}(H_k^{n_k}) = \hat{H}^n$.

In DOVER, this mapping is done incrementally by considering the hypotheses pair-wise, with their order decided based on the average DER to all other hypotheses. Suppose, for example, that the hypotheses have been ordered as $H_1,\ldots,H_K$. In this case, $H_1$ and $H_2$ are first mapped together using $H_1$ as reference to obtain $H_{1,2}$, which is then mapped together with $H_3$ to obtain $H_{1,2,3}$, and so on, until we have $H_{1,\ldots,K}$. At each iteration, the mapping is performed using the Hungarian method~\cite{Kuhn1955TheHM}, similar to how the reference and system outputs are mapped to a common space for evaluating diarization error rate (DER). This is an instance of the weighted bipartite graph matching problem, and is poly-time solvable using linear sum assignment on the weight matrix. As shown in Fig.~\ref{fig:hungarian}, the graph consists of speaker labels as vertices, and the edge weights $M_{ij}$ are computed as the overlap duration between the corresponding speaker labels $i$ and $j$.

This incremental approach for label mapping treats it as an incremental assignment problem~\cite{Toroslu2007IncrementalAP}. Due to this treatment, it is unable to use the global pair-wise costs to map all the hypotheses to a common space simultaneously. Furthermore, the method is dependent on the order of hypotheses $H_k$, and choosing a bad initial pair may be detrimental to the final mapping. In Section~\ref{sec:dl_mapping}, we propose a new label mapping algorithm that overcomes these limitations.

\begin{figure}[t]
\centering
\includegraphics[width=0.7\linewidth]{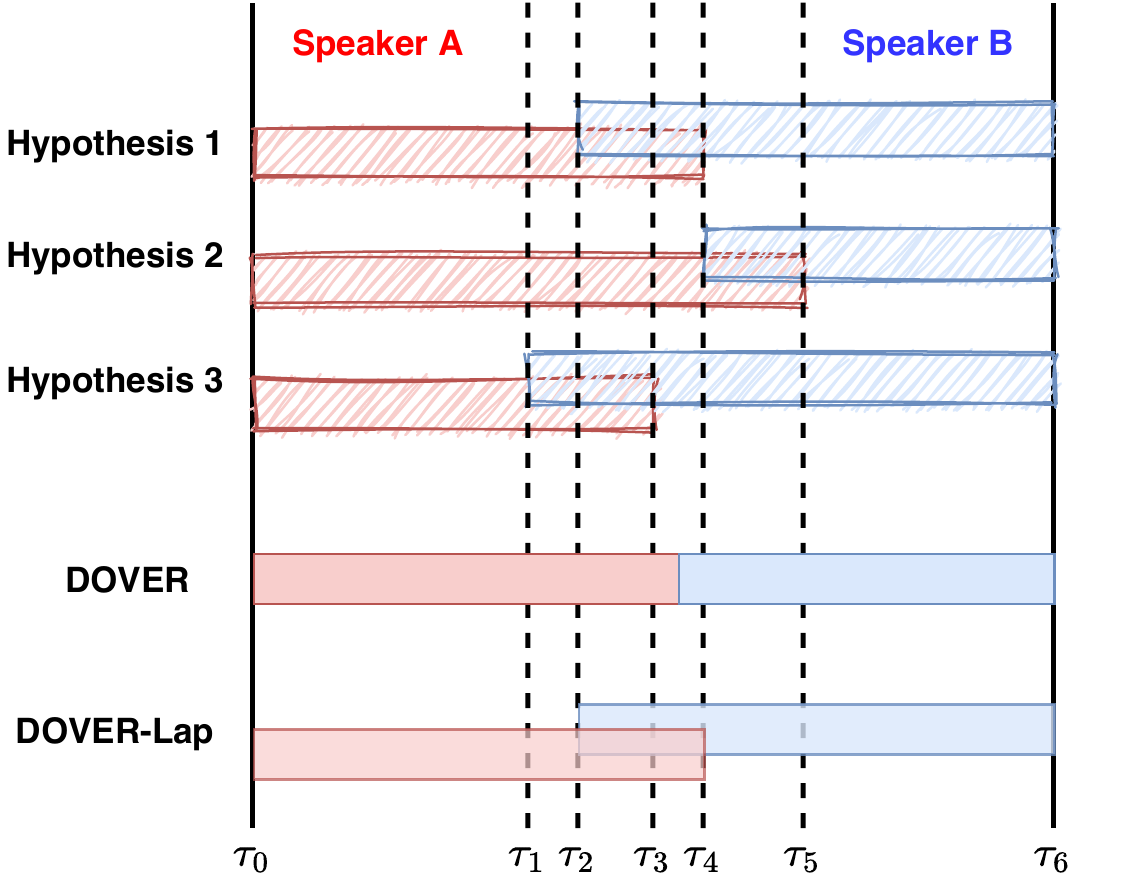}
\caption{An illustration of overlapping output produced by DOVER-Lap for overlapping hypotheses.}
\label{fig:dover_vs_dl}
\end{figure}

Once the hypotheses have been mapped to a common speaker label space, the \textbf{label voting} stage is performed. In DOVER, weighted majority voting is done on regions of input speech, where a region is defined as the maximal segment delimited by any of the original speaker boundaries from the input hypotheses (such as the intervals $[\tau_0,\tau_1],\ldots,[\tau_5,\tau_6]$ in Fig.~\ref{fig:dover_vs_dl}). Recall that overlap-aware diarization systems may contain multiple speakers in each such region. However, DOVER makes the single-speaker assumption in majority voting. Formally, if the audio is divided into $T$ regions, and $L_k = (l_1^k,\ldots,l_{\tau}^k,\ldots,l_T^k)$ denotes the region-wise labels assigned by hypothesis $k \in K$, where $l_{\tau}^k \in \{\hat{H}^1, \ldots, \hat{H}^N\}$, and $N$ is the number of speakers in combined label space. Then, the DOVER label voting stage computes, $\forall \tau \in T$,
\begin{equation}
\label{eqn:dover}
l_{\tau} = \arg \max_{\hat{H}^n \in \{\hat{H}^1, \ldots, \hat{H}^N\}} \left( \sum_{k \in K} w_k \mathbbm{1}(l_{\tau}^k = \hat{H}^n) \right),
\end{equation}
where $w_k \in [0,1]$ denotes a confidence weight assigned to hypothesis $k$. DOVER ranks the input hypotheses by their average DER to all other hypotheses, and applies a weight that decays slowly with rank: $w_k = \frac{1}{k^{0.1}}$. Since only a single speaker is assigned to every region, combination using DOVER may lead to high missed speech in the overlap case, as shown in Fig.~\ref{fig:dover_vs_dl}. We solve this problem through overlap-aware weighted majority voting, described in Section~\ref{sec:dl_voting}.

\subsection{DOVER-Lap: label mapping}
\label{sec:dl_mapping}

Given $K$ hypotheses, the problem of mapping them to a common label space is an instance of maximal weighted $K$-partite matching, which is known to be NP-hard for $K > 2$~\cite{Karp2010ReducibilityAC}. We propose an approximation algorithm for this problem, which is inspired from related work on 3-dimensional matching~\cite{Kann1991MaximumB3,Ausiello1999ComplexityAA}. While the DOVER label mapping technique solves the problem incrementally, our method considers all the pair-wise edge weights simultaneously, which leads to a better mapping.

\begin{figure}[t]
\centering
\includegraphics[width=0.9\linewidth]{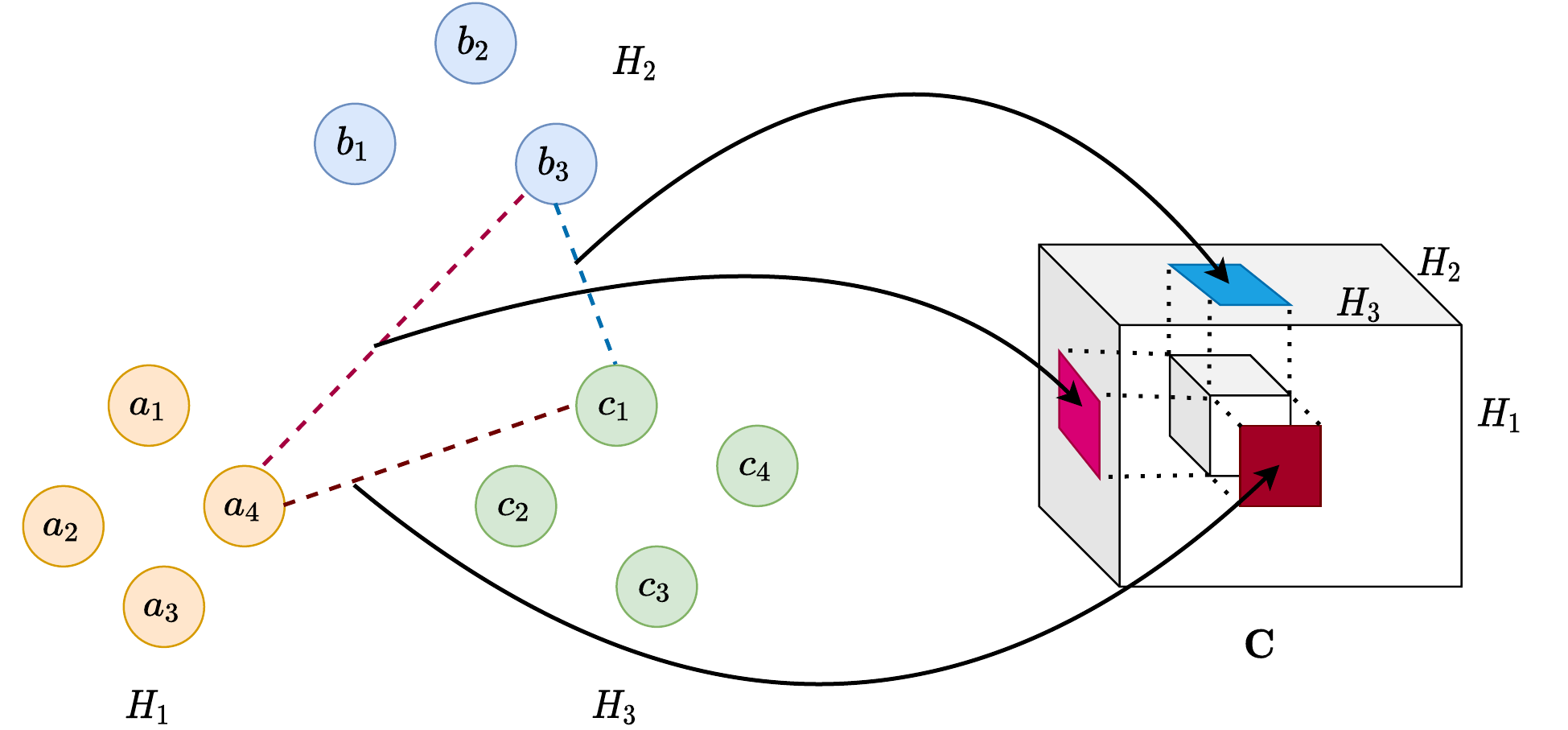}
\caption{Computation of the \textit{cost tensor} $\mathbf{C}$ for the DOVER-Lap label mapping algorithm.}
\label{fig:dl_mapping}
\end{figure}

Suppose hypothesis $H_k$ has $N_k$ speakers. We first compute a \textit{cost tensor} $\mathbf{C} \in \mathbb{R}^{N_1\times \ldots \times N_K}$, where each element of the tensor is defined as
\begin{align}
    C&(i_1,\ldots,i_K) = -\sum_{i=i_1}^{i_K} \sum_{j\succ i} M_{ij} \\
    &= - \scriptstyle{\left(M_{i_1,i_2}+\ldots+M_{i_1,i_K}\right) +\left(M_{i_2,i_3}+\ldots+M_{i_2,i_K}\right)} \nonumber \\
    &\scriptstyle{+ \ldots + \left(M_{i_{K-1},i_K}\right)}
\end{align}
where $M_{ij}$ is the overlap duration between speaker labels $i$ and $j$. Intuitively, $C(i_1, i_2,\ldots,i_K)$ computes the total cost if the speakers $i_1, i_2,\ldots,i_K$ from the respective hypotheses are mapped to the same label. This is obtained by adding all the $\binom{K}{2}$ edge costs. Since the hypotheses may contain overlapping speaker assignments, $M_{ij}$ computed here must be ``relative'' overlap ratios computed with respect to the total speaking time, otherwise a hypothesis which contains all the speakers for the entire duration will incur the lowest cost.

This computation is illustrated in Fig.~\ref{fig:dl_mapping} for $K=3$. Suppose the speaker labels in the hypotheses are denoted by the graph nodes as shown in this figure. Since the hypotheses have 4, 3, and 4 speakers, respectively, $C \in \mathbb{R}^{4\times 3\times 4}$. In the figure, we show the computation of the tensor index corresponding to the tuple $(a_4,b_3,c_1)$. For this, we first compute all 3 pair-wise costs; the cost of mapping $a_4$ and $b_3$ together, for instance, is computed as the negative of the edge weight, which is the total overlapping duration between $a_4$ and $b_3$, divided by the sum of their speaking durations. Finally, $C(a_4,b_3,c_1)=-(M_{a_4,b_3}+M_{a_4,c_1}+M_{b_3,c_1})$. 

With $\mathbf{C}$ thus computed, our objective is to find the minimal-weighted set of tuples $(i_1, i_2, \ldots, i_K)$ which span the entire tensor, i.e., which cover all the speaker labels. For this, we use a greedy approximation algorithm that iterates over the set of tuples constructing a maximal matching until all the speaker labels have been included in at least one tuple. We present the details of this method through pseudocode in Algorithm~\ref{alg:dl_mapping}. In line 3, we filter out those tuples for which all their indices (i.e., speaker labels) have been covered in the matching $\mathcal{M}$, and then sort them in increasing order of cost $C$. In lines 5--7, we iterate over the ordered list of tuples and add a tuple to $\mathcal{M}$ if it does not ``contradict'' $\mathcal{M}$, i.e., none of its labels are already present in some tuple of $\mathcal{M}$.  

\begin{algorithm}[t]
\DontPrintSemicolon
  
  \KwInput{Cost tensor $\mathbf{C}$}
  \KwOutput{Set of tuples $\mathcal{M}$ = \{$(i_1,\ldots,i_K)$\}}
  \KwInit{$R$ := \{$\bigcup_{n=1}^{N_k}(H_k,n)$: $k\in\{1,\ldots,K\}$\}} \tcp*{remaining labels}
 
  $\mathcal{M} = \phi$;
  
  \tcc{Loop until no labels are remaining}
  \While{$R$ is not empty}{
    
    \tcc{Get sorted list of tuples which have at least one label unmapped}
    $S$ = sorted(\{$(i_1,\ldots,i_K)$ if $\bigcup_k\left((H_k,i_k)\in R\right)$\}, key=$C(i_1,\ldots,i_K)$);
    
    \tcc{Find maximal matching}
    $\mathcal{M}^{\prime}=\phi$
    
    \For{tuple in $S$}{
        \tcc{Add tuple if none of its elements are present in the matching}
        \If{tuple not contradicts $\mathcal{M}^{\prime}$}
        {
            $\mathcal{M}^{\prime} = \mathcal{M}^{\prime} \cup \{tuple\}$
        }
    }
    
    $\mathcal{M} = \mathcal{M} \cup \mathcal{M}^{\prime}$
    
    $R = R \setminus \mathcal{M^{\prime}}$
  }

\caption{DOVER-Lap label mapping}
\label{alg:dl_mapping}
\end{algorithm}

\vspace{-0.8em}
\subsection{DOVER-Lap: label voting}
\label{sec:dl_voting}

Similar to DOVER, we perform weighted majority voting on ``regions'' of the input. However, unlike the former, DL can assign multiple speakers to the region. Consider a region $T$, and suppose the hypotheses $H_1,\ldots,H_K$ contain $n_1^T,\ldots,n_K^T$ speakers, respectively, in this region. Then, we compute the weighted mean rounded to the nearest integer as
\begin{equation}
\label{eq:dl_weight}
    \hat{n}_T = \nint{\sum_{k=1}^K w_k n_k^T},
\end{equation}
where $w_k$ are DOVER-like rank-based weights obtained by ranking the hypotheses in increasing order of their total relative overlap duration with all other hypotheses. The highest weighted $\hat{n}_T$ speakers are then assigned to the region $T$. In case of ties, we uniformly divide the region and assign each speaker to one of the sub-regions. Since we use rank-based weighting of the hypotheses, such ties only occur in very few regions. This assignment strategy allows multiple overlapping speakers to be present in the combined diarization output.

\subsection{DOVER vs. DOVER-Lap}

As with DOVER, DL also consists of a label mapping stage and a label voting stage, but modifies the algorithm in each of these stages. For the label mapping stage, DOVER uses the Hungarian method in a pair-wise incremental manner, while DL applies a global mapping strategy which considers all edge weights simultaneously. We will see in Section~\ref{sec:results} that this label mapping technique also improves DOVER for combining single-speaker hypotheses.

For the label voting stage, the core difference is illustrated in Fig.~\ref{fig:dover_vs_dl}. Suppose the hypotheses have been mapped to the same label space \{A,B\} in the label mapping stage. If we consider the region $[\tau_2,\tau_3]$, speaker A is present in all 3 hypotheses while speaker B is present in only 2 of them. As such, DOVER assigns speaker A to this region. DL, on the other hand, computes $\hat{n}_T = 2$ for this region, and assigns it to the top 2 speakers, which are A and B, in this case.

\vspace{0.5em}
\noindent
\textbf{Complexity analysis.} Suppose the $K$ hypotheses contain $N$ speakers each, on average. Since DOVER applies the Hungarian algorithm sequentially, the time complexity of its label mapping stage is $\mathcal{O}(K N^3)$. In DL, the cost tensor is computed in $\mathcal{O}(K^2 N^2)$ time, and the maximal matching algorithm has an amortized complexity of $\mathcal{O}(KN^K \log N)$ since we have to sort all $N^K$ tuples. In practice, our implementation of the algorithm ran in a few seconds on an i7 processor.

\vspace{-0.5em}
\section{System Combination Experiments}

\subsection{Dataset}

We performed experiments on two datasets: the AMI meeting corpus~\cite{Carletta2005TheAM}, and the LibriCSS data~\cite{Chen2020ContinuousSS}. AMI consists of 100 hours of recorded meetings containing 4 speakers per session, with speech from close-talking, single distant microphone (SDM), and array microphones. About 20\% of the speech duration contains overlaps, on average. For our experiments, we used the mixed-headset recordings, which are obtained by summing the individual headset signals from the participants in the meeting. LibriCSS is a recently released corpus consisting of multi-channel audio recordings of ``simulated conversations''. The simulated mixtures were created by combining utterances from the LibriSpeech corpus~\cite{Panayotov2015LibrispeechAA}. LibriCSS comprises 10 sessions, where each session is approximately one hour long. Each session is made up of six 10-minute-long ``mini sessions'' that have different overlap ratios, ranging from 0 to 40\%, and contains 8 speakers. The recordings were made in a regular meeting room by using a seven-channel circular microphone array. For our experiments, we selected the recordings from the first channel of the array. 

\vspace{-0.7em}
\subsection{Diarization methods}

In this paper, since our focus is on combining overlap-aware diarization outputs, we show our main results using the following diarization systems, which can assign overlapping speakers.

\begin{enumerate}[leftmargin=*]

\item \textbf{VB-based overlap assignment (VB)}~\cite{Bullock2019OverlapawareDR}: This method leverages Variational Bayes (VB)-HMM used originally for diarization in \cite{Dez2018SpeakerDB}. 
Using the output of an externally trained overlap detector, overlapping frames are assigned the top two speakers from the posterior matrix computed using VB inference. We used a single-speaker spectral clustering system to initialize the matrix with hard probabilities~\cite{Park2020AutoTuningSC}, and an oracle speech activity detector (SAD) to remove non-speech segments.

\item \textbf{Overlap-aware spectral clustering (SC)}~\cite{Raj2021MultiSC}: This method also uses an external overlap detector, but unlike VB, overlap assignment happens in the first-pass clustering itself, instead of during resegmentation. This is done by reformulating spectral clustering as a constrained optimization problem, and then discretizing it under the overlap constraints. We used the same SAD, x-vector extractor, and overlap detector for the VB and SC methods.

\item \textbf{Region proposal networks (RPN)}~\cite{Huang2020SpeakerDW}: It combines segmentation and embedding extraction into a single neural network, and jointly optimizes them using an objective function that consists of boundary prediction and speaker classification components. The region embeddings are then clustered (using K-means clustering) and a non-maximal suppression is applied. For AMI, we trained the RPN on force-aligned data from the AMI training set; for LibriCSS, it was trained on simulated meeting-style recordings with partial overlaps generated using utterances from the LibriSpeech~\cite{Panayotov2015LibrispeechAA} training set. Since we used K-means clustering, we assumed that the oracle number of speakers for each recording is known. A post-processing step was applied using oracle SAD segments to filter non-speech.

\item \textbf{Target-speaker voice activity detection (TS-VAD)}~\cite{Medennikov2020TargetSpeakerVA}: It takes conventional speech features (e.g., MFCC) along with i-vectors for each speaker as inputs and produces frame-level activities for each speaker using a neural network with a set of binary classification output layers. The initial estimates for the speaker i-vectors were obtained using a spectral clustering system. For training the model for LibriCSS (we did not use TS-VAD for the AMI experiments), we created simulated meeting-style data similar to that used for training the RPN model.
\end{enumerate}

\subsection{Diarization results on AMI}
\label{sec:results}

Table~\ref{tab:ami_results} shows the results on the AMI mix-headset data. We obtained diarization outputs using the VB, SC, and RPN models, and then combined them with DOVER and our proposed DL method. Since DOVER is not designed to handle overlapping speaker assignments, it showed 19.5\% and 19.9\% missed speech for the dev and test sets, respectively, which is equal to the overlap durations. Additionally, we found that the speaker confusion error was also higher than the average over the inputs (+4.5\% for dev and +1.3\% for eval), which we attribute to poor label mapping. We replaced the DOVER label mapping with our proposed global algorithm, and the speaker confusion rates improved significantly over the input average (-6.0\% for dev and -4.3\% for eval), although the missed speech still remained high. Consequently, the total DER for DOVER with global mapping was found to be comparable to the average over all inputs. This improvement was even more significant on the single-speaker regions, where the DER reduced by more than 50\% with the new label mapping algorithm. This suggests that our mapping algorithm can be effectively applied to combine system outputs even in low-overlap conditions.

Next, we applied DL to combine the hypotheses, with and without rank-based weighting (denoted by $w_k$ in Equation~\eqref{eq:dl_weight}). Even without rank weighting, DL was found to consistently improve over the average, from 27.3\% to 24.1\% for dev, and 23.5\% to 22.8\% for test, respectively. With rank-based weighting, the method outperformed the best single system for both evaluation sets. On further investigation, we found that this improvement came largely from reduced missed speech on the overlapping regions, with 40.7\% and 34.4\% relative reduction on the dev and test sets, respectively. This shows that estimating the number of speakers in a region through weighted average, as done in Equation~\eqref{eq:dl_weight}, is effective for label voting. On single-speaker regions, DL performs slightly worse than DOVER with global mapping since it may produce false alarms, while DL is constrained to predict at most one speaker.

\begin{table}[t]
\small
\centering
\caption{Comparison of our proposed DL method with overlap-aware diarization baselines and DOVER, in terms of DER (\%), on the AMI mix-headset data. The dev and test sets contain 19.5\% and 19.9\% overlapped speech, respectively. Numbers in smaller font denote DERs when scored on the single-speaker regions only.}
\label{tab:ami_results}
\begin{tabular}{@{}lcc@{}}
\toprule
\textbf{Method} & \textbf{Dev} & \textbf{Test} \\ \midrule
VB-based overlap assignment & 22.0 {\tiny 12.3} & 21.5 {\tiny 8.5} \\
Overlap-aware SC & 24.5 {\tiny 12.8} & 23.6 {\tiny 9.0} \\
Region proposal networks & 35.3 {\tiny 29.8} & 25.5 {\tiny 16.4} \\
Average & 27.3 {\tiny 18.3} & 23.5 {\tiny 11.3} \\
\midrule
DOVER & 36.5 {\tiny 20.4} & 30.5 {\tiny 11.2} \\
+ global mapping (proposed) & 26.0 {\tiny 7.7} & 25.0 {\tiny 5.2} \\
\midrule
DL (w/o rank weighting) & 24.1 {\tiny 9.6} & 22.8 {\tiny 6.6} \\
DL (with rank weighting) & \textbf{21.6} {\tiny 10.8} & \textbf{20.5} {\tiny 7.4} \\ \bottomrule
\end{tabular}
\end{table}

\begin{table}[t]
\centering
\caption{Diarization performance on LibriCSS evaluation set (sessions 2-10), evaluated condition-wise, in terms of \% DER. 0S and 0L refer to 0\% overlap with short and long inter-utterance silences, respectively. The DL results are using rank-based weighting.}
\label{tab:libricss}
\begin{adjustbox}{max width=\linewidth}
\begin{tabular}{@{}lccccccc@{}}
\toprule
\multicolumn{1}{c}{\multirow{2}{*}{\textbf{Method}}} & \multicolumn{6}{c}{\textbf{Overlap ratio in \%}} & \multicolumn{1}{c}{\multirow{2}{*}{\textbf{Average}}} \\
\cmidrule(r{4pt}){2-7}
\multicolumn{1}{c}{} & \multicolumn{1}{c}{\textbf{0L}} & \multicolumn{1}{c}{\textbf{0S}} & \multicolumn{1}{c}{\textbf{10}} & \multicolumn{1}{c}{\textbf{20}} & \multicolumn{1}{c}{\textbf{30}} & \multicolumn{1}{c}{\textbf{40}} & \multicolumn{1}{c}{} \\
\midrule
VB & 3.9 & 3.8 & 6.5 & 8.2 & 12.6 & 13.4 & 8.6 \\
SC & 2.6 & 3.4 & 6.8 & 10.0 & 13.9 & 15.2 & 9.3 \\
RPN & 4.5 & 9.1 & 8.3 & 6.7 & 11.6 & 14.2 & 9.5 \\
TS-VAD & 6.0 & 4.6 & 6.6 & 7.3 & 10.3 & 9.5 & 7.4 \\
\midrule
DL & \textbf{2.3} & \textbf{2.2} & \textbf{4.0} & \textbf{5.3} & \textbf{7.9} & \textbf{9.0} & \textbf{5.4} \\
\bottomrule
\end{tabular}
\end{adjustbox}
\end{table}

\vspace{-0.8em}
\subsection{Diarization results on LibriCSS}

Since LibriCSS does not have an official dev/eval split, we used the first session for development, and the remaining 9 sessions for evaluation. In Table~\ref{tab:libricss}, we show the DER results for the baseline diarization systems and DL, with a break down by overlap condition. We also report a further break down by missed speech, false alarm, and speaker confusion, in Table~\ref{tab:libri_breakdown}.

Similar to the results on AMI, we found that DL improved the average DER over the single best system (TS-VAD, in this case) from 7.4\% to 5.4\%. This improvement was consistent across the different overlap conditions, even though the single best system themselves may differ depending on the condition. Furthermore, from Table~\ref{tab:libri_breakdown}, we see that DL was able to combine the complementary strengths of the diarization systems. For instance, VB and SC performed better on overlap detection (as indicated by their lower MS and FA), while RPN and TS-VAD had lower speaker confusion. By combining these system outputs using DL, high speaker accuracy was obtained without worsening detection (MS and FA) rates.

\begin{table}[t]
\centering
\caption{Diarization result break-down on LibriCSS evaluation set, in terms of \% missed speech (MS), false alarm (FA), and speaker confusion (Conf.).}
\label{tab:libri_breakdown}
\begin{tabular}{@{}lcccc@{}}
\toprule
\textbf{Method} & \textbf{MS} & \textbf{FA} & \textbf{Conf.} & \textbf{DER} \\ \midrule
VB & \textbf{1.7} & \textbf{0.5} & 6.4 & 8.6 \\
SC & 2.5 & 1.1 & 5.7 & 9.3 \\
RPN & 2.9 & 3.3 & 3.3 & 9.5 \\
TS-VAD & 3.2 & 1.3 & 2.9 & 7.4 \\
\midrule
DL & 2.7 & 0.7 & \textbf{2.0} & \textbf{5.4} \\ \bottomrule
\end{tabular}
\end{table}

\vspace{-0.6em}
\section{Late Fusion for Multi-microphone Diarization}
\vspace{-0.6em}

The DOVER-Lap algorithm combines diarization hypotheses, irrespective of the source of these hypotheses. In the above, we have applied it for combining different diarization systems; however, it can also be applied to several other use cases. For instance, we may have a single-channel diarization system, but input signals from an array microphone. In such cases, the system can be independently run on each channel, and the outputs can be combined using DL --- this is a classic ``late fusion'' application (in contrast to early fusion techniques such as beamforming~\cite{HaebUmbach2019SpeechPF,Hain2007TheAS}). Another similar application of the algorithm is for diarization on ad-hoc arrays~\cite{Yoshioka2019MeetingTU}, where audio is captured on the participant's mobile devices, and this ``ad-hoc array'' is used for speech processing. 

In this section, we demonstrate the application of DL for late fusion on array microphones. We conducted our investigation on the LibriCSS dataset, which has 7 microphones arranged in a circular array. For our diarization system, we used the overlap-aware spectral clustering (SC)~\cite{Raj2021MultiSC} method described earlier. As shown in Table~\ref{tab:libri_breakdown}, the method obtained a DER of 9.3\% on LibriCSS using a single microphone.

Table~\ref{tab:fusion} shows the results for multichannel diarization using late fusion with DL. The single-channel system obtained a DER of 9.40\% on average (with a standard deviation of 0.23\%). For early fusion, we applied online weighted prediction error (WPE)~\cite{Nakatani2010SpeechDB} based dereverberation followed by delay-and-sum beamforming on the input channels. We used the Nara implementation~\cite{Drude2018NARAWPEAP} of WPE and the Beamformit tool~\cite{Mir2007AcousticBF} for beamforming. The corresponding DER was found to be 9.33\%, which is marginally better than the 7-channel average. Notably, simple beamforming without dereverberation degraded the DER to 9.71\%. Late fusion using DL improved over the average and best single system by achieving 9.02\% DER. Similar to our earlier results, we found that the improvement was mostly from reduced false alarms and speaker confusions.

\begin{table}[]
\centering
\caption{Diarization results for multichannel LibriCSS evaluation set. Late fusion using DL achieved better performance compared to early fusion based on dereverberation and beamforming.}
\label{tab:fusion}
\begin{adjustbox}{max width=\linewidth}
\begin{tabular}{@{}lcccc@{}}
\toprule
\textbf{Method} & \textbf{MS} & \textbf{FA} & \textbf{Conf.} & \textbf{DER} \\ \midrule
7-channel avg. & \textbf{2.58} & 0.96 & 5.86 & 9.40 \\
7-channel best & 2.59 & 0.99 & 5.53 & 9.11 \\
\midrule
WPE + Beamforming & 2.91 & 0.96 & 5.86 & 9.33 \\
DL & 3.60 & \textbf{0.66} & \textbf{4.76} & \textbf{9.02} \\ \bottomrule
\end{tabular}
\end{adjustbox}
\end{table}

\vspace{-0.5em}
\section{Conclusion}

We proposed a new method, DOVER-Lap (DL), to combine the outputs from overlap-aware diarization systems. Our method was inspired by the label mapping and label voting approach in DOVER, but modified the algorithms used in each of these stages. We replaced incremental pair-wise mapping with a global mapping strategy based on a cost tensor, and incorporated overlap awareness in label voting. We demonstrated through experiments on meeting datasets (AMI and LibriCSS) that DL is effective at combining the outputs from different kinds of diarization systems, such as clustering-based, RPN, and TS-VAD. It provided consistent and significant improvements over the single best system for both datasets. We also showed its applicability to multichannel diarization through late fusion, where it outperformed early fusion methods. Although DL is effective at combining overlap-aware diarization outputs, the label voting method may need to be modified if we have a mix of single-speaker and overlapping outputs. Additionally, there is scope for further improvement in label mapping, since the current method processes the cost tensor in a greedy manner, which may not be optimal. These investigations are left as future work.

{
\small
\textbf{Acknowledgments}. Some of the baselines reported here were implemented during JSALT 2020 at JHU, with support from Microsoft, Amazon, and Google. We thank Maokui He for providing the TS-VAD diarization output on LibriCSS.
}

\bibliographystyle{IEEEbib}

\small

\bibliography{mybib}

\end{document}